\documentclass[12pt]{article}
\usepackage{amssymb}
\usepackage{amsmath}

\newtheorem{theorem}{Theorem}

\newtheorem{corollary}{Corollary}

\begin{document}

\author{{Roustam Zalaletdinov} \\ [5mm]
\emph{Department of Mathematics and Statistics, Dalhousie University} \\
\emph{Chase Building, Halifax, Nova Scotia, Canada B3H 3J5} \\ [2mm]
\emph{Department of Theoretical Physics, Institute of Nuclear Physics} \\
\emph{Uzbek Academy of Sciences, Ulugbek, Tashkent 702132, Uzbekistan, CIS} \\
}
\title{{\LARGE \textbf{Averaging out Inhomogeneous Newtonian Cosmologies: I.
Fluid Mechanics and the Navier-Stokes Equation }}}
\date{}
\maketitle

\begin{abstract}
The basic concepts and equations of classical fluid mechanics are presented
in the form necessary for the formulation of Newtonian cosmology and for
derivation and analysis of a system of the averaged Navier-Stokes-Poisson
equations. A special attention is paid to the analytic formulation of the
definitions and equations of moving fluids and to their physical content.
\end{abstract}

\section{Introduction}

In looking for adequate gravitational field equations applicable in the
cosmological settings \cite{Elli:1984} there has been a number of attempts
to average Einstein equations (for a review and references see \cite
{Zala:1992}, \cite{Kras:1997} and recent papers \cite{Boer:1998}, \cite
{SHT:1999}). Despite four decades of research in this field since the
pioneering paper \cite{Shir-Fish:1962}, there is still no definite consensus
on the results and approaches to the averaging problem in general relativity
and cosmology. It does reflect to a certain extent the fundamental character
of this problem. The averaging problem in general relativity has actually
originated in cosmology \cite{Elli:1984}-\cite{Kras:1997}, \cite
{Shir-Fish:1962}-\cite{Tava-Zala:1998}, but the main questions remain
unanswered. A definite and unambiguous picture of the averaged dynamics of
the universe is not available as yet.

In view of the difficulties mentioned above, there have been some attempts
to construct an averaged theory of the Newtonian self-gravitating fluid to
describe on its basis various aspects of dynamics of homogeneous and
inhomogeneous Newtonian cosmologies, such as backreaction, large-scale
structure formation, scaling properties (for a review and references see
\cite{Olso-Sach:1973}, \cite{Yodz:1974} and recent papers \cite
{Sota-etal:1998}-\cite{TSMK:2001}). Studying the Newtonian gravity and
cosmology is of importance for three main reasons:

(i) The physics of Newtonian self-gravitating configurations provides an
adequate theoretical framework for modelling the structure formation, galaxy
dynamics, clustering processes, etc. in cosmology \cite{Peeb:1993};

(ii) It is well-known that Newtonian cosmology has very similar physics in
many aspects with the general relativistic cosmology \cite{Bond:1960}, \cite
{Trau:1965};

(iii) Newtonian problems are simpler to deal with than the general
relativistic ones and, given similarities and differences between them \cite
{MLE:1998}, \cite{Elli:2000}, one can gain important intuition and
experience to approach the similar general-relativistic cosmological
problems..

In order to derive a system of the averaged Navier-Stokes-Poisson equations
and to make use of the system for analysis of inhomogeneous Newtonian
universes, it is significant to present the basic concepts and equations of
classical fluid mechanics in the form necessary to approach this problem. In
this presentation a special attention will be paid to the analytic
formulation of the definitions and equations of moving fluids and their
physical content.

The structure of this paper is as follows. Chapter 2 gives a brief account
on the interconnection between fluid mechanics and Newtonian cosmology. In
Chapter 3 some approaches to averaging inhomogeneous Newtonian cosmologies
are reviewed and discussed. The main concepts regarding the fluid motion and
the Eulerian and Lagrangian pictures of moving fluids are presented in
Chapters 4. Chapter 5 presents the field description of moving fluids which
provides a framework for an analytic formulation of fluid motion as the
theory of the integrals of partial differential equations. The Euler
expansion formula for the change of the volume of an infinitesimal element
of fluid during its motion is proved in Chapter 6. The Reynolds transport
theorem following from the Euler expansion formula for the change of an
integral of a fluid quantity over an arbitrary fluid volume moving along the
fluid paths is proved in Chapter 7. In Chapter 8 the fluids possessing the
property of inertia are introduced and the equation of continuity for the
fluid density is derived as a consequence of the fluid mass conservation.
The Principle of conservation of linear momentum is formulated in Chapter 9
and the Cauchy equation of motion is obtained as a consequence of this
conservation. The structure of the system of the Cauchy equation of motion
and the equation of continuity for the description of fluid motion is also
analyzed here. The kinematic quantities to characterize the change of a
fluid element during its motion, namely, the expansion scalar, the shear
tensor and the vorticity vector, are introduced and discussed in Chapter 10.
Chapter 11 is devoted to analysis of the conservation of energy for fluids
and formulation of the laws of thermodynamics for fluids with a special
attention paid to the following energy transfer equations and the definition
of the thermodynamical variables for the compressible and incompressible
fluids. The next Chapter gives a discussion of the conservation of angular
momentum and its relation with the structure of the fluid stress tensor.
Chapter 13 provides a list of different forms of the Cauchy equation of
motion in dependence on the structure of fluids. In particular, the Euler
equation for the perfect fluid, the Stokes equation for the Stokean fluid
and the Navier-Stokes equation of classical hydrodynamics for the Newtonian
fluid are presented and discussed. The physical status of the Navier-Stokes
equation as the macroscopic equation governing the dynamics of gases and
liquids is also discussed.

Conventions and notations are as follows. All functions $f=f(x^{i},t)$ are
defined on 3-dimensional Euclidean space $E^{3}$ in the Cartesian
coordinates $\{x^{i}\}$ with Latin space indices $i,j,k,...$ running from 1
to 3, and $t$ is the time variable. The Levi-Civita symbol $\varepsilon
_{ijk}$ is defined as $\varepsilon _{123}=+1$ and $\varepsilon ^{123}=+1$
and $\delta ^{ij}$, $\delta _{j}^{i}$ and $\delta _{ij}$ are the Kronecker
symbols. The symmetrization of indices of a tensor $
T_{jk}^{i}=T_{jk}^{i}(x^{l},t)$ is denoted by round brackets, $T_{(jk)}^{i}=
\frac{1}{2}\left( T_{jk}^{i}+T_{kj}^{i}\right) $, and the antisymmetrization
by square brackets, $T_{[jk]}^{i}=\frac{1}{2}\left(
T_{jk}^{i}-T_{kj}^{i}\right) $. A partial derivative of $T_{jk}^{i}$ with
respect to a spatial coordinate $x^{i}$ or time $t$ is denoted either by
comma, or by the standard calculus notation, $T_{jk,t}^{i}=\partial
T_{jk}^{i}/\partial t$ and $T_{jk,l}^{i}=\partial T_{jk}^{i}/\partial x^{l}$
, the fluid velocity is $u^{i}=u^{i}(x^{j},t)$,\ and the material (total)
derivative is $\dot{T}_{jk}^{i}\equiv dT_{jk}^{i}/dt=\partial
T_{jk}^{i}/\partial t+u^{l}\partial T_{jk}^{i}/\partial
x^{l}=T_{jk,t}^{i}+u^{l}T_{jk,l}^{i}$.

\section{Fluid Mechanics and Newtonian Cosmology}

The dynamics of a compressible Newtonian fluid evolving in its own Newtonian
gravitational potential field is known to be governed by the
Navier-Stokes-Poisson system (see \cite{Moni-Yagl:1971}, \cite{Batc:1967}
and \cite{Elli:1971}, \cite{Elli:1973} for the system of equations in the
framework of the so-called 3+1 covariant formulation of general relativistic
cosmology and references therein). It consists of the equation of continuity
which states the conservation of mass of a fluid element, the Navier-Stokes
equation which is the Newton second law for the element of fluid stating the
conservation of the fluid element momentum, the Poisson equation for the
fluid gravitational potential and an equation of state of the fluid. The
fluid flow is described by its velocity, density, pressure, or temperature,
and the molecular transport coefficients of dynamic and bulk viscosity
and/or thermal conductivity. Generally, the evolution of a fluid element can
be characterized by change of its kinematic characteristics: the expansion
scalar of the fluid element characterizes the change of its volume, the
shear tensor determines the distortion of the fluid element leaving its
volume unchanged and the vorticity vector represents its rigid rotation. The
Navier-Stokes equation and the equation of continuity can be therefore
rewritten explicitly as a system of evolution equations for the expansion,
the shear, the vorticity and the fluid density. In case of Newtonian
cosmology the system of the Navier-Stokes-Poisson equations for a
compressible Newtonian fluid without viscosity and heat transfer is assumed
to describe the Newtonian universes, that is, possibly spatially infinite
distributions of the evolving self-gravitating cosmological fluid governed
by the Navier-Stokes-Poisson equations. For homogeneous and isotropic
Newtonian universes only time-dependent functions of the expansion, the
pressure and the density remain essential. The evolution equation for the
expansion scalar then describes the dynamics of evolving homogeneous and
isotropic Newtonian universes. This equation is known to have the same form
as the Raychaudhuri equation in general relativity \cite{Rayc:1955}, \cite
{Hawk:1966} describing the evolution of the expansion scalar of a
4-dimensional flow of the cosmological fluid in homogeneous and isotropic
general relativistic cosmology. It is equivalent to the Friedmann equation
\cite{Frie:1922} for a scale factor of an expanding universe in both
Newtonian and general relativistic cases. Therefore, studying Newtonian
cosmology proved its importance for understanding some aspects of general
relativistic cosmology and revealed remarkable similarities between both
theoretical approaches to studying the cosmological models of the real
Universe \cite{Bond:1960}, \cite{Trau:1965}.

In case of inhomogeneous and anisotropic Newtonian universes, when the whole
system of the equations must be considered, one can study the dynamics of a
Newtonian universe which is smooth, homogeneous and/or isotropic, on large
scales. In the general relativistic setting an inhomogeneous cosmological
model which has the Friedmann-Lema\^{\i}tre-Robertson-Walker (FLRW)
space-time geometry \cite{Frie:1922}-\cite{KSHM:1980} on average is
generally believed to be more realistic than the standard FLRW model of our
Universe \cite{Elli:1984}. An essential dynamical factor which must affect
the character of dynamics of such an inhomogeneous universe is that it takes
into account the effect of the backreaction of inhomogeneities on the global
expansion of the universe (for a review and references see \cite
{Bild-Futa:1991} and recent papers \cite{Taka-Futa:1999}, \cite{TSMK:2001},
\cite{Namb:2002}). In case of Newtonian cosmology a similar setting is also
of great interest for understanding the cosmological backreaction due to
correlations of a self-gravitating cosmological fluid. To construct such\ a
model a suitable averaging procedure must be applied to the
Navier-Stokes-Poisson system, to bring about a set of equations for the
averaged expansion, shear, vorticity, pressure, fluid density and
gravitational potential together with their correlation functions involved.

The nonrelativistic kinematics and dynamics of fluid flows have been
intensively studied since the eighteenth century (for a review and
references see \cite{Serr:1959}, \cite{True-Toup:1960}), which resulted in
discovering the Navier-Stokes equation as the fundamental equation of
classical hydrodynamics \cite{Navi:1827}, \cite{Stok:1845}. This equation
successfully describes the motion of various fluids and gases and other
transport phenomena under a wide class of physical conditions including
viscosity and heat\ transfer \cite{Batc:1967}. The existence of two
different types of flow, called laminar and turbulent, has been established
in the middle of the nineteenth century. The pioneering systematic
investigations of turbulence by Rayleigh and Reynolds in the last decades of
the nineteenth century led to understanding that the correlation functions
of the fluid velocity and the fluid density must be recognized as the
fundamental characteristics of turbulence (for a review and references see
\cite{Moni-Yagl:1971}, \cite{Moni-Yagl:1975}-\cite{Stan:1985}). The first
system of differential equations for the correlation functions containing
the single-point first and second moments of the fluid fields has been
derived by Reynolds \cite{Reyn:1894} using the time and space averagings of
the system of the continuity and the Navier-Stokes equations for the case of
incompressible fluids. An analytical formulation of the problem of
turbulence has been given first by Keller and Friedmann \cite{Kell-Frie:1924}
who developed a general method for obtaining the differential equations for
the moments of arbitrary order by using the system of the continuity and the
Navier-Stokes equations. The Reynolds equations are the first equations in
the total infinite system of the Keller-Friedmann equations. Analysis of
these systems of equations has resulted in remarkable progress in our
understanding of the phenomenon of turbulence to establish the statistical
nature of turbulence, to develop the theory of homogeneous and isotropic
turbulence, to propose the hypothesis of locally isotropic turbulence and
developed turbulence \cite{Moni-Yagl:1971}, \cite{Moni-Yagl:1975}-\cite
{Stan:1985}. However, despite the significant developments the problem of
turbulence remains essentially unsolved since even such fundamental
questions as which flow should be considered turbulent and which should not,
what is a mechanism of the turbulent mixing, how the developed turbulence
occurs, etc. do not have satisfactory explanation as yet. The difficulties
originate partly in the mathematics of the Navier-Stokes and Reynolds
equations where the questions of existence, uniqueness, and smoothness of
solutions are not mainly resolved \cite{FMRT:2001}. On the other hand,
theoretical approaches to formulate adequate physical models of real
physical phenomena exhibiting turbulent processes are often very limited due
to the absence of sufficient experimental evidence about a turbulent regime
of interest \cite{Gibs:1996}.

\section{Approaches to Averaging out Inhomogeneous Universes}

As it has been pointed out above, the physics of Newtonian self-gravitating
fluid configurations is widely employed for modelling cosmological
gravitating structures and the Universe a whole. The idea of Chaos, or
Cosmic Turbulence, as the origin of our Universe goes back to Kepler,
Descartes and Laplace \cite{Jone-Peeb:1972}. Development of the hot Big Bang
Model and attempts to understand the origin of galaxies has led to invoking
the hydrodynamic turbulence theory in modelling the galaxy formation \cite
{vonW:1947}, \cite{Gamo:1952} as opposed to the idea of the origin of
galaxies due to the gravitational instability (see \cite{Peeb:1993}, \cite
{Zel'-Novi:1974} for discussion and references). It has been shown \cite
{Ozer-Cher:1968}, in particular, that the physical conditions in the early
universe could allow the Primeval Turbulence regime and the appearance of
eddies to produce galaxies later on.

However, application of the Navier-Stokes-Poisson system and its suitably
averaged version for analysis of the evolution of inhomogeneous and
anisotropic Newtonian universes has not been accomplished in any
considerable depth. Such important questions as backreaction, large-scale
structure formation, scaling properties, dynamics of universe which is
homogeneous and isotropic on average and other dynamical issues related to
the character of the cosmological evolution when the gravitational
correlation effects are taken into account are still open. In this setting
the system of equations for the 3+1 covariant formulation of general
relativistic cosmology has been hardly studied either.

A study of the averaged-over-ensemble self-gravitating Newtonian perfect
fluid in the Eulerian formulation has been carried out first in \cite
{Olso-Sach:1973} under assumption of the homogeneous isotropic turbulence
\cite{Batc:1953}. Despite this strong requirement which leads to a
considerable simplification of the system of the averaged
Navier-Stokes-Poisson equations, the Newtonian cosmological models exhibit
nontrivial dynamics. The evolution equation for the mean-square correlation
function of vorticity has been derived and its solution has been shown to
have a nontrivial behavior to blow up in\ a finite time for an expanding
background in case of domination of the fragmentation of eddies
(inhomogeneities), and to blow up always in a finite time for a contracting
background. Thus, in dependence of a value of the initial vorticity the
averaged evolution of an initially inhomogeneous Newtonian cosmological
model can develop either a turbulence mixing regime, or a regime of the
natural decay of inhomogeneities.

Analysis of the general relativistic Raychaudhuri equation has been
attempted in \cite{Yodz:1974} for the case of closed universes which are
globally splitted into a topological product of compact, orientable,
differentiable 3-manifolds and the real line which represents time. These
3-manifolds can be naturally represented as a family of disjoint spacelike
hypersurfaces parametrized by time, and there exists a future-oriented
normal unit 4-vector field tangent to a normal congruence of curves which
are orthogonal trajectories of the hypersurface family. A general
relativistic analogue of the Reynolds transport theorem of fluid mechanics
\cite{Serr:1959}, \cite{True-Toup:1960}, \cite{Reyn:1903} for the rate of
change of the integral of the expansion scalar of this 4-vector flow over a
hypersurface moving with the flow has been derived. A consequent formula for
the rate of change of the ratio of this integral to the volume of the
corresponding hypersurface, which has been called the \textquotedblleft
average expansion\textquotedblright , was used to find out\ if this
\textquotedblleft average expansion\textquotedblright\ has a zero volume
singularity characteristic to the FLRW models. As a result of this analysis
it has been suggested that such a singularly can be avoided at all if during
some epochs the universe is expanding highly nonuniformly.

This result, however, should not be misinterpreted. Since by its kinematic
meaning a transport formula for a quantity is essentially an integral
formulation of the corresponding differential law for this quantity, there
is no new either kinematic, or dynamical information which can be gained by
using such an integral formulation. So, for example, the conservation of the
mass of fluid in a material volume moving with the fluid leads to the
equation of continuity which is well-known to be a necessary and sufficient
condition for a motion to conserve the mass of an arbitrary moving volume
\cite{Serr:1959}. On the contrary, the averaged continuity equation obtained
by the application of a properly defined space, time or ensemble average and
the corresponding formulae for interchanging of operations of taking a
derivative and an averaging is known to be a new differential dynamical
equation, the continuity equation for the turbulent flow \cite
{Moni-Yagl:1971}, \cite{Stan:1985}. Therefore, the conclusions derived in
\cite{Yodz:1974} on the basis of using a ratio of the integral of the
expansion scalar over an arbitrary hypersurface moving along the normal
4-vector flow to the volume of this region as a space average of the
expansion and using the transport formula for this ratio as a formula for
interchanging of the time derivative and this \textquotedblleft
averaging\textquotedblright\ cannot be considered as a consistent averaging
procedure. There is no new information in the derived integral formula as
compared with the corresponding local differential relation which is the
general relativistic Raychaudhuri equation.

Some recent approaches to construct an averaged theory of self-gravitating
Newtonian fluid to study the dynamical properties of the homogeneous and
inhomogeneous Newtonian universes, such as backreaction, large-scale
structure formation, scaling properties (see \cite{Sota-etal:1998}-\cite
{TSMK:2001} and references therein), have also made use of the similar ratio
described above \cite{Yodz:1974} as a \textquotedblleft space
average\textquotedblright\ of the expansion scalar and other kinematic
quantities for the case of 3-dimensional cosmological fluid. A
nonrelativistic version of the transport formula as in \cite{Yodz:1974}
which is known from the Reynolds transport theorem \cite{Serr:1959}, \cite
{True-Toup:1960}, \cite{Reyn:1903} modified by division by the region volume
has been taken as a formula for interchanging of the time derivative and
this \textquotedblleft averaging\textquotedblright\ and a set of the
transport relations for the kinematic quantities has been derived. This set
of transport relations was regarded as the averaged equations for the
kinematic quantities. In particular, the transport equation for the
expansion scalar has been taken as the averaged Newtonian version of
Raychaudhuri equation to analyze the character of the averaged\ dynamics.
Again, such a procedure of space averaging is not proper and all conclusions
are achievable by analyzing the initial system of the differential equations
for kinematic quantities.

\section{The Eulerian and Lagrangian Pictures of Fluid Motion}

\label{*elpfm}

The material fluid is considered to be a continuous physical substance
distributed over a region of the 3-dimensional Euclidean space, which is
assumed to possess a positive volume. A fluid flow, or a fluid motion, is
represented mathematically by a one-parameter family of mappings of a
3-dimensional region filled with the fluid into itself, the mapping being
usually assumed to be 1-1 and at least thrice continuously differentiable in
all variables \cite{Serr:1959}, \cite{True-Toup:1960}, \cite{True:1954},
\cite{Aris:1962} except possibly at certain singular surfaces, curves or
points where a separate analysis is required. The real parameter $t$
describing the mappings of fluid regions is identified with the time with
its domain supposed to be $-\infty <t<+\infty $ where $t=0$ is an arbitrary
initial instant.

In order to describe the fluid motion analytically let us introduce a fixed
rectangular coordinate system $\{x^{i}\}$, each coordinate triple of which
is referred to as a position of an infinitesimal element of a fluid
distribution, or configuration, often called a \textquotedblleft fluid
particle\textquotedblright . If one observes such a fluid particle moving
with the fluid, it is easy to detect that the particle which was in the
position $\{\xi ^{i}\}$ has moved to position $\{x^{i}\}$ at a later time.
Without loss of generality, one can take the first instance to be $t=0$ and
if the later time is $t$ the fluid flow is represented by the transformation
\begin{equation}
x^{i}=x^{i}(\xi ^{j},t).  \label{flow}
\end{equation}
If $\xi ^{i}$ is fixed while $t$ varies, Eq. (\ref{flow}) specifies the path
of the fluid particle
\begin{equation}
x^{i}=x^{i}(t)\equiv x^{i}(\xi ^{j},t)_{\mid \xi ^{i}=\mathrm{const}},
\hspace{0.4cm}x^{i}(0)=\xi ^{i},  \label{path}
\end{equation}
initially placed at $\xi ^{i}$. If time $t$ is fixed, Eq. (\ref{flow})
determines a transformation
\begin{equation}
x^{i}=x^{i}(\xi ^{j})\equiv x^{i}(\xi ^{j},t)_{\mid t=\mathrm{const}},
\hspace{0.4cm}x^{i}(0)=\xi ^{i},  \label{region}
\end{equation}
of the region initially occupied by the fluid into its position at time $t$.
The transformation (\ref{flow}) is precisely the above-mentioned
one-parameter family of mappings defining the fluid motion
\begin{equation}
x^{i}:x^{i}(0)=\xi ^{i}\rightarrow x^{i}(\xi ^{j},t).  \label{flow-map}
\end{equation}
Since functions $x^{i}(\xi ^{j},t)$ are single-valued and thrice
differentiable in all variables, the Jacobian of the transformation (\ref
{flow}) does not vanish,
\begin{equation}
J=\frac{\partial (x^{1},x^{2},x^{3})}{\partial (\xi ^{1},\xi ^{2},\xi ^{3})}
=\det \left( \frac{\partial x^{i}}{\partial \xi ^{j}}\right) \neq 0,
\label{jacobian}
\end{equation}
and there exists a set of inverse functions
\begin{equation}
\xi ^{j}=\xi ^{j}(x^{i},t)  \label{inverse-flow}
\end{equation}
to define the initial positions of the fluid particle which is at any
position $\{x^{i}\}$ at a moment of time $t$. This inverse transformation is
also 1-1 and thrice differentiable in all variables. Due to Eq. (\ref
{inverse-flow}) a fluid flow can be equivalently described by a set of
initial positions of its particles in dependence of their positions in later
times.

Thus, there are two equivalent ways to describe the fluid motion.\vspace{
0.2cm}

\noindent \textbf{The Eulerian description by the spatial variables }$
(x^{i},t)$\hspace{0.2cm}\emph{This picture presents the state of motion at
each position }$\{x^{i}\}$\emph{\ such that throughout all time a given set
of coordinates }$\{x^{i}\}$\emph{\ remains attached to a fixed position,
serving as an identification for it - the spatial variables are called the
Eulerian or spatial coordinates.}\vspace{0.2cm}

\noindent \textbf{The Lagrangian description by the material variables }$
(\xi ^{i},t)$\hspace{0.2cm}\emph{This picture of the motion chronicles the
history of each fluid particle }$\{\xi ^{i}\}$\emph{\ such that throughout
all time a given set of coordinates }$\{\xi ^{i}\}$\emph{\ remains attached
to a single fluid particle, serving to uniquely identify it - the material
variable are called the Lagrangian, material or convected coordinates}
\footnote{
Though the descriptions of the fluid motion by the spatial and material
coordinates are usually called Eulerian and Lagrangian, correspondingly,
both of them are known to be due to Euler \cite{Serr:1959}, \cite{True:1954}.
}.\vspace{0.2cm}

It should be pointed out here that the Eulerian and Lagrangian descriptions
have quite different physical interpretations when one would like to
describe a moving fluid from the point of view of an observer carrying out
measurements by a set of physical instruments. In the Eulerian picture an
observer is always located in a particular position $\{x^{i}\}$ at a moment
of time $t$ and such an observer views moving fluid particles passing
through and by the position. In the Lagrangian picture an observer is always
moving with a fluid particle which was initially at a position $\{\xi ^{i}\}$
and such an observer views the relative changes in the fluid comoving with
the observer's particle.

\section{The Field Description of Fluid Motion}

\label{*fdfm}

Although the fluid flow is completely determined either by the
transformation (\ref{flow}) in the Eulerian description, or its inverse
transformation (\ref{inverse-flow}) in the Lagrangian description, it is
important also to be able to describe the state of motion at a given
position during the course of time. To achieve this goal one should
characterize the motion of fluid in terms of objects defined at a position $
\{x^{i}\}$ at a time $t$. A simplest object to be considered for this role
is the instant fluid velocity $u^{i}$, functions of which can be defined in
both descriptions, $u^{i}(x^{j},t)$ and $u^{i}(\xi ^{j},t)$.

Let us consider a quantity $f$ which characterizes a fluid in the framework
of the Eulerian or Lagrangian descriptions. Generally, any such a quantity $
f $ is a function of spatial variables $f=f(x^{i},t)$ and it is also a
function of material variables $f=f(\xi ^{i},t)$ due to
\begin{equation}
f=f[x^{i}(\xi ^{j},t),t]\hspace{0.4cm}\mathrm{or}\hspace{0.4cm}f=f[\xi
^{i}(x^{j},t),t],  \label{field}
\end{equation}
the functions (\ref{field}) being related by Eqs. (\ref{flow}) and (\ref
{inverse-flow}) considered as the laws of the change of the variables.
Geometrically, $f=f(x^{i},t)$ is the value of $f$ felt by the fluid
particles instantaneously located at the position $\{x^{i}\}$ and $f=f(\xi
^{i},t)$ is the value of $f$ experienced at time $t$ by the fluid particle
initially located at the position $\{\xi ^{i}\}$. Then the change in
quantity $f$ when the fluid moves can be characterized by the two different
time derivatives
\begin{equation}
\frac{\partial f}{\partial t}\equiv \frac{\partial f(x^{i},t)}{\partial t}
_{\mid x^{i}=\mathrm{const}}\hspace{0.4cm}\mathrm{and}\hspace{0.4cm}\frac{df
}{dt}\equiv \frac{\partial f(\xi ^{i},t)}{\partial t}_{\mid \xi ^{i}=\mathrm{
const}}.  \label{time-drs}
\end{equation}
The partial derivative $\partial f/\partial t$ gives the rate of change of $
f $ with respect to a fixed position $\{x^{i}\}$, while the material, or
convective, derivative $df/dt$ measures the rate of change of $f$ with
respect to a moving fluid particle.

The material derivative of the position of a fluid particle defined by the
second Eq. (\ref{time-drs}) for $f=x^{i}$ is called its velocity $u^{i}$
\begin{equation}
u^{i}(\xi ^{j},t)=\frac{dx^{i}}{dt}\equiv \frac{\partial x^{i}(\xi ^{j},t)}{
\partial t}_{\mid \xi ^{i}=\mathrm{const}}.  \label{velocity}
\end{equation}
The notion of the fluid velocity plays a fundamental role in the field
formulation of fluid mechanics. Indeed, in most physical settings, when the
actual motion of a fluid particle, that is either $x^{i}=x^{i}(\xi ^{j},t)$
or $\xi ^{j}=\xi ^{j}(x^{i},t)$, is not given explicitly, it is more
convenient to deal with the Eulerian form of the fluid velocity, $
u^{i}=u^{i}(x^{j},t)$. Given a vector field of fluid velocity $
u^{i}=u^{i}(x^{j},t)$, which can be actually measured in laboratory, defined
in a fluid 3-region $U$ for a time interval $\Delta t$, one can determine
the functions $x^{i}=x^{i}(\xi ^{j},t)$ by solving the initial value problem
for the system of the first order ordinary differential equations
\begin{equation}
\frac{dx^{i}}{dt}=u^{i}(x^{j},t),\quad x^{i}(0)=\xi ^{i}.
\label{velocity-eqs}
\end{equation}
A unique solution $x^{i}=x^{i}(\xi ^{j},t)$ to Eqs. (\ref{velocity-eqs}) is
known (see, for instance, \cite{Birk-Rota:1989}, \cite{NSS:2000}) to exist
always on $U\times \Delta t$ for at least once continuously differentiable
in all variables and bounded functions $u^{i}=u^{i}(x^{j},t)$, and it
depends continuously on the initial values $\{\xi ^{i}\}$, to completely
characterize the fluid motion (\ref{flow}). Therefore, the description of
fluid motion in terms of the fluid velocity is equivalent to the Eulerian or
Lagrangian pictures:\vspace{0.2cm}

\noindent \textbf{The field description by the fluid velocity in the spatial
variables }$u^{i}(x^{j},t)$\hspace{0.2cm}\emph{This picture presents the
state of a fluid in motion by its velocity field specifying the velocity }$
u^{i}=u^{i}(x^{j},t)$\emph{\ of a fluid particle at each position }$
\{x^{i}\} $\emph{\ at each time instance }$t$,\emph{\ such that throughout
all time a given set of velocity components }$\{u^{i}\}$\emph{\ remains
attached to a fixed position, serving as an identification for a single
fluid particle in dependence of its initial position }$\{\xi ^{i}\}$\emph{\
- the velocity variables are called the field variables}\footnote{
Euler and d'Alembert are considered to be the first to discover the field
description of moving fluids. However, it was Euler who put forward the idea
of formulating a field theory of fluid motion as the theory of the integrals
of partial differential equations \cite{Serr:1959}, \cite{True:1954}.}.
\vspace{0.2cm}

The physical interpretation of the field description of fluid motion
represents an observer as located in a particular position $\{x^{i}\}$ at a
moment of time $t$ to measure the velocities of the fluid particles moving
by the observer and which were initially at $\{\xi ^{i}\}$.

The field description of moving fluid has a fundamental significance for the
proper analytic treatment of processes occurring in physical continuous
media under a wide class of physical conditions. It is the field description
that provides the possibility of an analytic formulation of fluid kinematics
and dynamics by finding a set of partial differential equations governing
the fluid motion.

With the definitions of the time derivatives (\ref{time-drs}) one can
calculate the material derivative of any quantity $f(x^{j},t)$ as
\begin{eqnarray}
\frac{df}{dt} &=&\left[ \frac{\partial }{\partial t}f(\xi ^{i},t)\right]
_{\mid \xi ^{i}=\mathrm{const}}=  \notag \\
&&\left\{ \frac{\partial }{\partial t}f[x^{i}(\xi ^{j},t),t]\right\} _{\mid
\xi ^{i}=\mathrm{const}}=\frac{\partial f}{\partial t}_{\mid x^{i}=\mathrm{
const}}+\frac{\partial f}{\partial x^{i}}\left[ \frac{\partial x^{i}(\xi
^{j},t)}{\partial t}\right] _{\mid \xi ^{i}=\mathrm{const}}
\label{material-dr1}
\end{eqnarray}
that can be written in the following form by using the fluid velocity $
u^{i}=u^{i}(x^{j},t)$ defined by Eq. (\ref{velocity}):
\begin{equation}
\frac{df}{dt}=\frac{\partial f}{\partial t}+u^{i}\frac{\partial f}{\partial
x^{i}}.  \label{material-dr2}
\end{equation}
This formula relates the material and spatial derivatives through the
velocity field in the field picture of fluid motion and it expresses the
rate of change in $f(x^{j},t)$ with respect to a moving fluid particle
located at a position $\{x^{i}\}$ at time $t$. An important application of
the material derivative (\ref{material-dr2}) is the definition of the fluid
acceleration $a^{i}=a^{i}(x^{j},t)$,
\begin{equation}
a^{i}=\frac{du^{i}}{dt}=\frac{\partial u^{i}}{\partial t}+u^{j}\frac{
\partial u^{i}}{\partial x^{j}},  \label{acceleration}
\end{equation}
as the rate of change of the velocity of fluid particle experienced during
its motion. Due to (\ref{acceleration}) the fluid acceleration $a^{i}$ is
represented only in terms of the fluid velocity $u^{i}$.

\section{The Euler Expansion Formula}

\label{*eef}

The Jacobian $J$ (\ref{jacobian}) of the transformation (\ref{flow})
represents the dilatation of the infinitesimal volume of a fluid particle as
it moves along its path (\ref{path}). Indeed, considering (\ref{flow}) as
the coordinate transformation from the initial coordinates $\{\xi ^{i}\}$ at
time $t=0$ of a fluid particle to the particle coordinates $\{x^{i}\}$ at
later time $t$, the change of the particle's infinitesimal volume is given
by $dV=JdV_{0}$ where $dV=dx^{1}dx^{2}dx^{3}$ is the particle volume at the
position $\{x^{i}\}$ and $dV_{0}=d\xi ^{1}d\xi ^{2}d\xi ^{3}$ is the initial
particle volume. Because the fluid motion is continuous and (\ref{flow}) is
continuously differentiable and therefore invertible, the fluid dilatation,
or expansion, $dV/dV_{0}$, is positive and bounded
\begin{equation}
J=\frac{dV}{dV_{0}},\quad 0<J<\infty .  \label{dilatation}
\end{equation}
Since the infinitesimal volume of a fluid particle undergoes, in general,
changes and distortions during its motion, the rate of change of the
dilatation characterizes such changes. The formula for the dilatation change
had been found by Euler \cite{Eule:1755} (see \cite{Serr:1959}, \cite
{True-Toup:1960}, \cite{True:1954}, \cite{Aris:1962} for discussion).

\begin{theorem}[The Euler expansion formula]
The rate of change of the fluid dilatation $J$ (\ref{dilatation}) following
a fluid particle path is given by
\begin{equation}
\frac{dJ}{dt}=J\frac{\partial u^{i}}{\partial x^{i}},\hspace{0.4cm}\mathrm{or
}\hspace{0.4cm}\frac{d}{dt}\ln J=\frac{\partial u^{i}}{\partial x^{i}}.
\label{euler}
\end{equation}
\end{theorem}

\noindent \textbf{Proof.}\hspace{0.2cm}A derivation of the Euler formula is
straightforward by differentiation of the Jacobian determinant, $
dJ=J\partial \xi ^{i}/\partial x^{j}d\left( \partial x^{i}/\partial \xi
^{j}\right) $,
\begin{equation}
\frac{dJ}{dt}=J\frac{\partial \xi ^{j}}{\partial x^{i}}\frac{d}{dt}\left(
\frac{\partial x^{i}}{\partial \xi ^{j}}\right) =J\frac{\partial \xi ^{j}}{
\partial x^{i}}\frac{\partial u^{i}}{\partial \xi ^{j}}=J\frac{\partial \xi
^{j}}{\partial x^{i}}\frac{\partial u^{i}}{\partial x^{k}}\frac{\partial
x^{k}}{\partial \xi ^{j}}=J\frac{\partial u^{i}}{\partial x^{i}}.
\label{euler-proof}
\end{equation}
\textbf{QED}\vspace{0.2cm}

Since the Euler expansion formula (\ref{euler}) establishes the divergence
of the fluid velocity as a measure of the fluid dilatation, or expansion,
the former is also called the fluid expansion and denoted as $\theta
=\partial u^{i}/\partial x^{i}$. The Euler formula (\ref{euler}) emphasizes
the fundamental role of the fluid velocity in the field description of fluid
motion. When a fluid is moving without change in volume for all fluid
particles at all position $\{x^{i}\}$ throughout all time $t$,
\begin{equation}
\frac{dJ}{dt}=0\quad \Leftrightarrow \quad \frac{\partial u^{i}}{\partial
x^{i}}=0,  \label{incompressible}
\end{equation}
such a fluid is called incompressible. If a fluid is not incompressible,
\begin{equation}
\frac{dJ}{dt}\neq 0\quad \Leftrightarrow \quad \frac{\partial u^{i}}{
\partial x^{i}}\neq 0,  \label{compressible}
\end{equation}
is called compressible.

\section{The Transport Theorem of Reynolds}

\label{*ttr}

An important kinematical theorem follows from the Euler expansion formula
(\ref{euler}). It is called the Reynolds transport theorem \cite{Reyn:1903}
and it concerns the rate of change not an infinitesimal volume of a fluid
particle, but any integral of any quantity $f$ which is a function of
spatial variables $f=f(x^{i},t)$ over an arbitrary closed fluid region, or
volume, $\Sigma (t)$ moving with the fluid \cite{Serr:1959}, \cite
{True-Toup:1960}, \cite{True:1954}, \cite{Aris:1962}. Let us consider a
volume integral of $f(x^{i},t)$ over $\Sigma (t)$,
\begin{equation}
F(t)=\int\limits_{\Sigma (t)}f(x^{i},t)dV,  \label{integral}
\end{equation}
which is a well-defined function $F=F(t)$ of time.

\begin{theorem}[The Reynolds transport theorem]
The rate of change of the volume integral (\ref{integral}) following the
fluid particle paths is given by
\begin{equation}
\frac{d}{dt}\int\limits_{\Sigma (t)}fdV=\int\limits_{\Sigma (t)}\left( \frac{
df}{dt}+f\frac{\partial u^{k}}{\partial x^{k}}\right) dV.
\label{reynolds-fml}
\end{equation}
\end{theorem}

\noindent \textbf{Proof.}\hspace{0.4cm}It is straightforward by
transformation of (\ref{integral}) into the integral over the fixed initial
volume $\Sigma _{0}=\Sigma (0)$ in the material coordinates $\{\xi ^{i}\}$
\begin{equation}
\frac{d}{dt}F(t)=\frac{d}{dt}\int\limits_{\Sigma _{0}}f[x^{i}(\xi
^{j},t),t]JdV_{0}=\int\limits_{\Sigma _{0}}\left\{ \frac{d}{dt}f[x^{i}(\xi
^{j},t),t]J+f[x^{i}(\xi ^{j},t),t]\frac{dJ}{dt}\right\} dV_{0}
\label{reynolds-proof}
\end{equation}
which immediately brings (\ref{reynolds-fml}) by applying the Euler
expansion formula (\ref{euler}).\hspace{0.4cm}\textbf{QED}\vspace{0.2cm}

By substitution (\ref{material-dr2}) for the material derivative $df/dt$ in
the integrand of (\ref{reynolds-fml}), one can obtain the Reynolds transport
theorem in the following form
\begin{equation}
\frac{d}{dt}\int\limits_{\Sigma (t)}fdV=\int\limits_{\Sigma (t)}\left( \frac{
\partial f}{\partial t}+\frac{\partial }{\partial x^{k}}\left( fu^{k}\right)
\right) dV.  \label{reynolds-fml2}
\end{equation}
There is yet another useful form of the last formula obtained by application
of the Gauss theorem
\begin{equation}
\frac{d}{dt}\int\limits_{\Sigma (t)}fdV=\int\limits_{\Sigma (t)}\frac{
\partial f}{\partial t}dV+\int\limits_{\partial \Sigma (t)}fu^{k}dS_{k}
\label{reynolds-fml3}
\end{equation}
where the 2-surface element $dS_{k}$ is a vector equal to the infinitesimal
2-square and normal to the surface $\sigma (t)=\partial \Sigma (t)$
enclosing the volume $\Sigma (t)$. The formula (\ref{reynolds-fml3}) admits
an immediate physical interpretation as expressing the rate of change of a
quantity $f$ within a moving fluid volume as the integral of the rate of
change in this quantity itself at a point plus the net flow of $f$ over the
surface bounding the volume.

One particular case of the Reynolds transport theorem is of importance. For
the constant quantity $f=\mathrm{const}$ the Reynolds transport theorem (\ref
{reynolds-fml}) gives a formula for the change of value of the volume $
V_{\Sigma }(t)$ of an arbitrary moving fluid region $\Sigma (t)$,
\begin{equation}
V_{\Sigma }(t)=\int\limits_{\Sigma (t)}dV.  \label{volume-def}
\end{equation}

\begin{corollary}[The change of volume]
The rate of change of the value of the volume $V_{\Sigma }(t)$ (\ref
{volume-def}) of an arbitrary moving fluid region $\Sigma (t)$ is determined
by the formula
\begin{equation}
\frac{d}{dt}V_{\Sigma }(t)=\int\limits_{\Sigma (t)}\frac{\partial u^{k}}{
\partial x^{k}}dV=\int\limits_{\partial \Sigma (t)}u^{k}dS_{k}.
\label{reynolds-vol}
\end{equation}
\end{corollary}

On the basis of the Euler formulas \cite{True:1954}, \cite{Eule:1760} for
the transformation of the infinitesimal line element of a fluid particle
path between a spatial line element $dx^{i}$ and a material line element $
d\xi ^{i}$,
\begin{equation}
dx^{i}=d\xi ^{k}\frac{\partial }{\partial \xi ^{k}}x^{i}(\xi ^{j},t)\hspace{
0.4cm}\mathrm{and}\hspace{0.4cm}d\xi ^{i}=dx^{k}\frac{\partial }{\partial
x^{k}}\xi ^{i}(x^{j},t),  \label{euler-arc}
\end{equation}
and the Nanson formulas \cite{True:1954}, \cite{Nans:1878} for the
transformation of the infinitesimal surface elements for a 2-dimensional
surface made of fluid particle paths between a spatial surface element $
dS_{i}$ and a material surface element $dS_{(0)i}$,
\begin{equation}
dS_{k}\frac{\partial }{\partial \xi ^{i}}x^{k}(\xi ^{j},t)=JdS_{(0)i}\quad
\mathrm{and}\quad dS_{(0)k}\frac{\partial }{\partial x^{i}}\xi
^{k}(x^{j},t)=JdS_{i},  \label{nanson-sur}
\end{equation}
one can derive the corresponding line and surface integral transport
relations \cite{True-Toup:1960}, \cite{True:1954}, \cite{Aris:1962} similar
to the Reynolds transport theorem.

The Reynolds transport theorem is an integral statement essentially
equivalent to the Euler expansion formula. The same true for the
corresponding line and surface integral theorems and the Euler and the
Nanson formulae (\ref{euler-arc}), (\ref{nanson-sur}).

As it has been discussed in Introduction, the Reynolds transport theorem has
been recently utilized for a definition of a space averaging over fluid
volumes. This issue will be discussed in \cite{Zala-Cole:2002}.

\section{The Equation of Continuity}

\label{*ec}

The definition of the fluid velocity (\ref{velocity}), the system of partial
differential equations (\ref{velocity-eqs}) for determination of the
equation of motion of fluid $x^{i}=x^{i}(\xi ^{j},t)$, the Euler expansion
formula (\ref{euler}) and the Reynolds transport theorem (\ref{reynolds-fml}
) constitute the kinematics of moving fluid. The formulation of kinematics
does not require any hypotheses about the structure and the nature of fluids
as physical substances. To establish the differential equations governing
the fluid dynamics, one must formulate such physical axioms on the basis of
experimental evidence.

As it has been mentioned above, in the framework of the field description
one must characterize the motion of fluids in terms of objects defined at a
position $\{x^{i}\}$ at a time $t$. The fluid velocity $u^{i}$ has been
shown to be the simplest quantity to formulate the kinematics of fluid. Let
us now further consider fluids which possess the property of inertia. Such a
fluid can be endowed with density function $\rho =\rho (x^{i},t)$ of the
physical dimension [mass/volume], which enables one to define the mass $M(t)$
of the fluid held within the region $\Sigma (t)$
\begin{equation}
M(t)=\int\limits_{\Sigma (t)}\rho dV.  \label{mass-reg}
\end{equation}
The fluid mass (\ref{mass-reg}) for any region $\Sigma (t)$ is assumed to be
always positive that means positivity of the density function,
\begin{equation}
\rho (x^{i},t)>0,  \label{mas-positive}
\end{equation}
everywhere. The following physical principle is taken to hold for the fluid.
\vspace{0.2cm}

\noindent \textbf{The Principle of \ conservation of mass}\hspace{0.2cm}
\emph{The mass of fluid }$M(t)$\emph{\ }(\ref{mass-reg})\emph{\ in an
arbitrary region }$\Sigma (t)$\emph{\ does not change following the fluid
particle paths}
\begin{equation}
\frac{d}{dt}M(t)=0\hspace{0.4cm}\mathrm{or}\hspace{0.4cm}\frac{d}{dt}
\int\limits_{\Sigma (t)}\rho dV=0.  \label{mass-conservation}
\end{equation}
\vspace{0.2cm}

The Principle of conservation of mass immediately brings about a
differential relation for the density function $\rho (x^{i},t)$\ that is
called the equation of continuity \cite{Serr:1959}, \cite{True-Toup:1960},
\cite{True:1954}, \cite{Aris:1962} discovered first by Euler \cite{Eule:1755}
.

\begin{theorem}[The equation of continuity]
The equation of continuity
\begin{equation}
\frac{d\rho }{dt}+\rho \frac{\partial u^{j}}{\partial x^{j}}=0
\label{eq-continuity}
\end{equation}
is necessary and sufficient for the mass of any region following the fluid
particle paths to be conserved (\ref{mass-conservation}).
\end{theorem}

\noindent \textbf{Proof.}\hspace{0.4cm}Using the Reynolds transport theorem
(\ref{reynolds-fml}) for the fluid mass $M(t)$ (\ref{mass-reg}) gives the
rate of change of the mass as
\begin{equation}
\frac{d}{dt}M(t)=\int\limits_{\Sigma (t)}\left[ \frac{d\rho }{dt}+\rho \frac{
\partial u^{k}}{\partial x^{k}}\right] dV.  \label{reynolds-mass}
\end{equation}
Necessity follows immediately upon substitution of the equation of
continuity (\ref{eq-continuity}) into (\ref{reynolds-mass}). Sufficiency
follows from the mass conservation (\ref{mass-conservation}) and (\ref
{reynolds-mass}) by taking into account that the fluid region $\Sigma (t)$
is arbitrary.\hspace{0.4cm}\textbf{QED}\vspace{0.2cm}

Using the formula for the material derivative (\ref{material-dr2}) gives
another form of the equation of continuity
\begin{equation}
\frac{\partial \rho }{\partial t}+\frac{\partial }{\partial x^{j}}(\rho
u^{j})=0.  \label{eq-continuity2}
\end{equation}

An important formula for the rate of change of an integral of a product of a
quantity $f(x^{i},t)$ by the density function $\rho (x^{i},t)$
\begin{equation}
\lbrack \rho f](t)=\int\limits_{\Sigma (t)}\rho (x^{i},t)f(x^{j},t)dV.
\label{product}
\end{equation}
is due to the Reynolds transport theorem (\ref{reynolds-fml}) and the
equation of continuity (\ref{eq-continuity}).

\begin{corollary}
For any quantity $f(x^{i},t)$ the rate of change of the volume integral of
the product $\rho (x^{i},t)f(x^{j},t)$ (\ref{product}) following fluid
particle paths is given by
\begin{equation}
\frac{d}{dt}\int\limits_{\Sigma (t)}\rho fdV=\int\limits_{\Sigma (t)}\rho
\frac{df}{dt}dV.  \label{eq-continuity-gen}
\end{equation}
\end{corollary}

The equation of continuity holds for any fluid satisfying the Principle of
conservation of mass (\ref{mass-conservation}). It should be pointed out
here that though the fluid mass is not a kinematic quantity the derivation
of the equation of continuity is fully due to the fluid kinematics. Given a
fluid distribution with the density $\rho (x^{i},t)$, one can consider a
problem of determination of the velocity field $u^{i}(x^{j},t)$ for the
fluid motion which preserves the mass of its each region. In such a setting
the equation of continuity (\ref{eq-continuity}) is the first order partial
differential equations to be solved to find the functions of the fluid
velocity. This equation is not sufficient in general to find 3 unknown
functions of the fluid velocity components $u^{i}(x^{j},t)$. The dynamics of
fluids cannot be therefore reconstructed only on the basis of the equation
of continuity and the fluid density cannot serve as the physical reason for
fluid motion.

\section{The Cauchy Equation of Fluid Motion}

\label{*cefm}

The fluid dynamics considers the physical forces, external and internal, to
act upon fluid particles to cause their motion, and,\ as a result, a motion
of the fluid. The forces acting on a moving fluid particle may be of two
kinds. Internal, or contact, forces are those acting on a fluid particle
through its bounding surface, for example, such as fluid pressure. External,
or body, forces are those acting throughout a medium, for example, such as
gravity or the electromagnetic force. Due to the Cauchy stress principle
\cite{Cauc:1827} over any closed surface $\partial \Sigma (t)$ there exists
a distribution of a stress vector $s^{i}=s^{i}(x^{j},t;n^{k})$ depending on
the normal vector $n^{k}$ to the surface, whose action and moment are
equivalent to the actual internal forces exerted by the fluid outside the
region enclosed by this surface upon the fluid inside \cite{Serr:1959}, \cite
{Aris:1962}, \cite{True:1952}. Such internal forces per unit surface are
represented by a vector $s_{(n)}^{i}=s_{(n)}^{i}(x^{j},t)$ directed along an
outward normal to the fluid particle surface. The external forces per unit
mass are represented by a vector $F^{i}=F^{i}(x^{j},t)$ which is assumed to
be known as functions of position $\{x^{i}\}$ and time $t$ and perhaps also
the state of the motion of the fluid. In accordance with the Cauchy stress
principle the following dynamical principle relates the linear momentum $
P^{i}(t)$ of an arbitrary fluid region $\Sigma (t)$
\begin{equation}
P^{i}(t)=\int\limits_{\Sigma (t)}\rho u^{i}dV  \label{momentum}
\end{equation}
with the corporate action of the stress vector $s_{(n)}^{i}$ and the
external forces $F^{i}$.\vspace{0.2cm}

\noindent \textbf{The Principle of conservation of linear momentum}\hspace{
0.2cm}\emph{The rate of change of linear momentum }$P^{i}(t)$ (\ref{momentum}
)\emph{\ of an arbitrary fluid region }$\Sigma (t)$\emph{\ is equal to the
total resultant force acting on the fluid volume within the region}\textsc{\
}
\begin{equation}
\frac{d}{dt}P^{i}(t)=\frac{d}{dt}\int\limits_{\Sigma (t)}\rho
u^{i}dV=\int\limits_{\Sigma (t)}\rho F^{i}dV+\oint\limits_{\partial \Sigma
(t)}s_{(n)}^{i}dS.  \label{momentum-conservation}
\end{equation}
\vspace{0.2cm}

By means of the formula (\ref{eq-continuity-gen}) the Principle of
conservation of linear momentum can be written in the following form:\textsc{
\ }
\begin{equation}
\int\limits_{\Sigma (t)}\rho \frac{du^{i}}{dt}dV=\int\limits_{\Sigma
(t)}\rho F^{i}dV+\oint\limits_{\partial \Sigma (t)}s_{(n)}^{i}dS.
\label{momentum-conservation2}
\end{equation}
In the representation (\ref{momentum-conservation2}) an arbitrary moving
region $\Sigma (t)$ can be replaced, without loss of generality, by a fixed
region $\Sigma (t_{0})$. An important theorem regarding the stress vector $
s^{i}$ holds \cite{Serr:1959}, \cite{True-Toup:1960}, \cite{Aris:1962}.

\begin{theorem}[The local equilibrium of the stress forces]
For any fluid moving with the conservation of linear momentum (\ref
{momentum-conservation}) the stress forces are in the local equilibrium
\begin{equation}
\lim_{\Sigma (t_{0})\rightarrow 0}\frac{1}{D^{2}}\oint\limits_{\partial
\Sigma (t_{0})}s_{(n)}^{i}dS=0.  \label{equilibrium}
\end{equation}
Here $D$ is the diameter of the region $\Sigma (t_{0})$. The volume $
V_{\Sigma }(t_{0})$ of the region $\Sigma (t_{0})$ is $V_{\Sigma
}(t_{0})\simeq D^{3}$ and the square $S_{\partial \Sigma }(t_{0})$ of the
region surface $\partial \Sigma (t_{0})\ $is $S_{\partial \Sigma
}(t_{0})\simeq D^{2}$.
\end{theorem}

\noindent \textbf{Proof.}\hspace{0.4cm}This result is derived by division of
both sides of (\ref{momentum-conservation2}) by $D^{2}$ and taking limit of $
\Sigma (t_{0})\rightarrow 0$. Then due to the boundedness of both integrands
for the volume integrals the formula (\ref{momentum-conservation2}) gives (
\ref{equilibrium}).\hspace{0.4cm}\textbf{QED}\vspace{0.2cm}

Analysis of the structure of the stress vector on the basis of its local
equilibrium (\ref{equilibrium}) gives the following result \cite{Serr:1959},
\cite{True-Toup:1960}, \cite{Aris:1962}, \cite{Cauc:1827}.

\begin{corollary}[The Cauchy formula for the stress vector]
The stress vector $s_{(n)}^{i}=s_{(n)}^{i}(x^{j},t)$ is a linear function of
the normal vector $n^{i}=n^{i}(x^{j},t)$ to the surface $\partial \Sigma (t)$
\begin{equation}
s_{(n)}^{i}=T^{i}{}_{j}n^{j}.  \label{stress-cauchy}
\end{equation}
Here the coefficients of the representation (\ref{stress-cauchy}) form a
tensor $T^{ij}=T^{ij}(x^{k},t)$ which is called the fluid stress tensor.
\end{corollary}

In general the second rank fluid stress tensor $T^{ij}(x^{k},t)$ has no
algebraic symmetry in its indices \cite{Serr:1959}, \cite{Hame:1927}.

As a result of the Cauchy formula (\ref{stress-cauchy}) one derives the
Cauchy equation of motion \cite{Serr:1959}, \cite{True-Toup:1960}, \cite
{Aris:1962}, \cite{Cauc:1827}.

\begin{theorem}[The Cauchy equation of motion]
The Cauchy equation of motion
\begin{equation}
\rho \frac{du^{i}}{dt}=\rho F^{i}+\frac{\partial T^{ij}}{\partial x^{j}}
\label{eqs-motion-cauchy}
\end{equation}
is necessary and sufficient for the linear momentum of any region following
the fluid particle paths to be conserved.
\end{theorem}

\noindent \textbf{Proof.}\hspace{0.4cm}The equation (\ref{eqs-motion-cauchy})
follows from the observation that the Cauchy formula for the stress tensor
(\ref{stress-cauchy}) and the Gauss theorem enable one to rewrite the
Principle of conservation of linear momentum (\ref{momentum-conservation2})
as \textsc{\ }
\begin{equation}
\int\limits_{\Sigma (t)}\rho \frac{du^{i}}{dt}dV=\int\limits_{\Sigma
(t)}\left( \rho F^{i}+\frac{\partial T^{ik}}{\partial x^{k}}\right) dV.
\label{momentum-conservation3}
\end{equation}
Now necessity follows immediately upon substitution of the equation of
motion (\ref{eqs-motion-cauchy}) into (\ref{momentum-conservation3}).
Sufficiency can be easily proved from the conservation of linear momentum
(\ref{momentum-conservation3}) by taking into account that the fluid region $
\Sigma (t)$ is arbitrary.\hspace{0.4cm}\textbf{QED}\vspace{0.2cm}

The Cauchy equation of motion is the most general equation valid for any
fluid motion which preserves the linear momentum of its each region
regardless of the form of the fluid stress tensor. From the point of view of
Newtonian mechanics the Cauchy equation of motion is the Newton second law
of motion stating that the acceleration $a^{i}(x^{j},t)$ (\ref{acceleration}
) of a fluid particle with the mass $\rho dV$ is due to the resulting force
of an external force $\rho F^{i}dV$ and an internal force $(\partial
T^{ij}/\partial x^{j})dV$ acting on the particle. In such a setting the
equation of motion (\ref{eqs-motion-cauchy}) is a system of the first order
partial differential equations to be solved to find the functions of the
components of the fluid velocity $u^{i}(x^{j},t)$. The equation involves the
density function $\rho (x^{i},t)$ which satisfies the equation of continuity
(\ref{eq-continuity}). Thus, \vspace{0.2cm}

\noindent \textbf{The Cauchy equation of motion for the fluid velocity }$
u^{i}$
\begin{equation}
\rho \left( \frac{\partial u^{i}}{\partial t}+u^{j}\frac{\partial u^{i}}{
\partial x^{j}}\right) =\rho F^{i}+\frac{\partial T^{ij}}{\partial x^{j}},
\label{eq1-cauchy}
\end{equation}
\vspace{0.2cm}

\noindent \textbf{The equation of continuity for the fluid density }$\rho $
\begin{equation}
\frac{\partial \rho }{\partial t}+\frac{\partial }{\partial x^{j}}(\rho
u^{j})=0,  \label{eq2-continuity}
\end{equation}
\vspace{0.2cm}

\noindent can be considered as a system of four first order partial differential
equations for four unknown functions of the components of the fluid velocity
$u^{i}(x^{j},t)$ and the fluid density $\rho (x^{i},t)$ provided the
external forces $F^{i}(x^{j},t)$ and the fluid stress tensor $
T^{ij}(x^{k},t) $ are given together with the initial and boundary
conditions for the velocity and density. The dynamics of a fluid is expected
to be reconstructed on the basis of solving this system. Upon finding the
fluid velocity $u^{i}(x^{j},t)$, the actual equation motion of the fluid
particles, either $x^{i}=x^{i}(\xi ^{j},t)$, or $\xi ^{j}=\xi ^{j}(x^{i},t)$
, is to be found by solving (\ref{velocity-eqs}).

The Cauchy equation of motion (\ref{eqs-motion-cauchy}) or (\ref{eq1-cauchy}
) can be written in another form by introduction of the total acceleration
vector $A^{i}=A^{i}(x^{j},t)$ which represents the combined effect of the
inertial force acting on a fluid particle through its acceleration $
a^{i}=a^{i}(x^{j},t)$ (\ref{acceleration}) and the effect of the external
force $F^{i}$,
\begin{equation}
A^{i}=\frac{du^{i}}{dt}-F^{i}\equiv a^{i}-F^{i}.  \label{acceleration-total}
\end{equation}
The Cauchy equation of motion then reads
\begin{equation}
\rho A^{i}=\frac{\partial T^{ij}}{\partial x^{j}}
\label{eqs-motion-cauchy-2}
\end{equation}
which is transparent from the physical point of view stating that the total
acceleration of a fluid particle is determined by action of the internal
force on the fluid particle.

\section{The Kinematics of a Moving Fluid Particle}

\label{*kmfp}

A fluid particle moving in accordance with the equations (\ref{eq1-cauchy})
and (\ref{eq2-continuity}) is subjected to changes in its shape, the value
of its volume and it generally rotates with respect to its initial, or any
other positions $\left\{ x^{i}\right\} $ at a time $t$, taken as a reference
point. Since it is the motion of a fluid particle under influence of the
internal and external forces (\ref{eq1-cauchy}) that affects its state, such
changes are due to its changing velocity in dependence of a position and a
moment of time. Let us consider the tensor $\partial u_{i}/\partial x^{j}$
constructed from the fluid velocity vector $u_{i}=\delta _{ij}u^{j}$, which
describes the spatial variations in the velocity of a fluid particle. This
is a second rank tensor has no algebraic symmetry in its indices and
therefore it can be generally decomposed into the following form:
\begin{equation}
\frac{\partial u_{i}}{\partial x^{j}}=\theta _{ij}+\omega _{ij},
\label{decomposition}
\end{equation}
where the symmetric tensor $\theta _{ij}=\theta _{ij}(x^{k},t)$ called the
deformation or rate of strain tensor is defined as
\begin{equation}
\theta _{ij}=\frac{1}{2}\left( \frac{\partial u_{i}}{\partial x^{j}}+\frac{
\partial u_{j}}{\partial x^{i}}\right) ,\hspace{0.4cm}\theta _{ij}=\theta
_{ji},  \label{deformation}
\end{equation}
and the antisymmetric tensor $\omega _{ij}=\omega _{ij}(x^{k},t)$ called the
vorticity tensor is defined as
\begin{equation}
\omega _{ij}=\frac{1}{2}\left( \frac{\partial u_{i}}{\partial x^{j}}-\frac{
\partial u_{j}}{\partial x^{i}}\right) ,\hspace{0.4cm}\omega _{ij}=-\omega
_{ji}.  \label{vorticity}
\end{equation}
In many case it is convenient to use the vorticity vector $\omega ^{i}$,
\begin{equation}
\omega ^{i}=\frac{1}{2}\varepsilon ^{ijk}\omega _{jk},  \label{vec-vorticity}
\end{equation}
instead of the vorticity tensor $\omega _{ij}$. It is easy to show that $
\omega _{ij}\omega ^{j}=0$.

The geometrical meaning of the deformation and the vorticity tensors are
given by the following theorem \cite{Serr:1959}, \cite{True-Toup:1960}, \cite
{Aris:1962} due to Cauchy \cite{Cauc:1841} and Stokes \cite{Stok:1845}.

\begin{theorem}[The Cauchy-Stokes decomposition theorem]
An arbitrary instantaneous state of a fluid particle moving along its path
may be resolved at each position $\{x_{(P)}^{i}\}$ as a superposition of a
uniform velocity of translation $u_{i(P)}=u_{i}(x_{(P)}^{k},t)$, a
dilatation $\partial \left[ x^{i}x^{j}\theta _{(P)ij}\right] /\partial x^{k}$
along three mutually perpendicular axes and\ a rigid rotation $\varepsilon
_{ijk}x^{j}\omega _{(P)}^{k}$ of these axes,
\begin{equation}
u_{i}(x^{k},t)=u_{i(P)}+\frac{1}{2}\frac{\partial }{\partial x^{i}}\left[
x^{j}x^{k}\theta _{(P)jk}\right] +\frac{1}{2}\varepsilon _{ijk}x^{j}\omega
_{(P)}^{k}+\mathcal{O}(x^{2}).  \label{cauchy-stokes}
\end{equation}
Here $\theta _{(P)ij}=\theta _{ij}(x_{(P)}^{k},t)$ is the deformation tensor
at $\{x_{(P)}^{i}\}$, $\omega _{(P)}^{k}=\omega _{(P)}^{k}(x_{(P)}^{k},t)$
is the vorticity vector at $\{x_{(P)}^{i}\}$.
\end{theorem}

\noindent \textbf{Proof.}\hspace{0.4cm}Let us consider the velocity field $
u^{i}(x^{k},t)$ of a moving fluid particle in a neighborhood of its position
$\{x_{(P)}^{i}\}$ at a time $t$. Since the velocity field is continuous and
differentiable, one has a Taylor expansion of the velocity functions near $
\{x_{(P)}^{i}\}$ which takes the following form with using the decomposition
(\ref{decomposition})
\begin{equation}
u_{i}(x^{k},t)=u_{i(P)}+\theta _{(P)ik}x^{k}+\omega _{(P)ik}x^{k}+\mathcal{O}
(x^{2}).  \label{velocity-p}
\end{equation}
The second term in (\ref{velocity-p}) can be written as
\begin{equation}
\theta _{(P)ik}x^{k}=\frac{1}{2}\frac{\partial }{\partial x^{i}}\left[
x^{j}x^{k}\theta _{(P)jk}\right] .  \label{deformation-p}
\end{equation}
This term represents a velocity field normal at each point to the quadric
surface $x^{j}x^{k}\theta _{(P)jk}=\mathrm{const}$ which contains the point $
\{x_{(P)}^{i}\}$. Since a symmetric tensor also possesses three mutually
perpendicular eigenvectors, the eigenvalues of the deformation tensor $
\theta _{ij}$ measure the rates of extension per unit length of a fluid
particle at $\{x_{(P)}^{i}\}$ in directions of the eigenvectors which can be
always taken as three basis vectors for the velocity field
(\ref{deformation-p}). By replacing the vorticity tensor by its vector
(\ref{vec-vorticity}) the last term in (\ref{velocity-p}) it takes the form
\begin{equation}
\omega _{(P)ik}x^{k}=\frac{1}{2}\varepsilon _{ijk}x^{j}\omega _{(P)}^{k}
\label{vorticity-p}
\end{equation}
with its immediate interpretation as representing a rigid rotation with an
angular velocity $\omega _{(P)}^{k}/2$.\hspace{0.4cm}\textbf{QED}\vspace{
0.2cm}

Thus, the deformation tensor $\theta _{ij}$ determines the distortion of the
element of fluid occupied by a fluid particle along the principal axes
defined by the eigenvectors of the tensor $\theta _{ij}$, namely a change in
the fluid particle's volume and shape. The vorticity tensor $\omega _{ij}$,
or vector $\omega ^{i}$, determines a rigid rotation of a fluid particle
with keeping its volume and shape the same. The rotation axis is defined by
the direction of the vorticity vector $\omega ^{i}$.

There are two important results following from the Cauchy-Stokes theorem.

\begin{corollary}[The rate of change of the fluid path arc]
The rate of change in the squared element of arc $dx^{i}$, $ds^{2}=\delta
_{ik}dx^{i}dx^{k}$, following a fluid particle path is given
\begin{equation}
\frac{d}{dt}(ds^{2})=2\theta _{ik}dx^{i}dx^{k}.  \label{deformation-arc}
\end{equation}
\end{corollary}

\begin{corollary}[The rigid motion]
For the rigid motion of a fluid when
\begin{equation}
\theta _{ij}(x^{k},t)=0  \label{rigid}
\end{equation}
everywhere, the fluid velocity $u^{i}(x^{k},t)$ is determined everywhere by
\begin{equation}
u^{i}=\frac{1}{2}\varepsilon _{ijk}x^{j}\omega ^{k}+\mathrm{const}.
\label{velocity-rigid}
\end{equation}
\end{corollary}

It is convenient to work sometimes with the trace-free deformation tensor
which is called the shear tensor $\sigma _{ij}=\sigma _{ij}(x^{k},t)$,
\begin{equation}
\sigma _{ij}=\theta _{ij}-\frac{1}{3}\delta _{ij}\theta ,\hspace{0.4cm}
\delta ^{ij}\sigma _{ij}=0,  \label{shear}
\end{equation}
where $\theta =\theta (x^{k},t)$ is the\ scalar of expansion,
\begin{equation}
\theta =\delta ^{ij}\theta _{ij}\equiv \frac{\partial u^{i}}{\partial x^{i}}.
\label{expansion}
\end{equation}
The shear tensor determines the distortion of a fluid particle along the
principal axes defined by its eigenvalues leaving the volume of the particle
unchanged. The magnitude of shear $\sigma $ is determined by $\sigma =\mid
\sigma ^{2}\mid ^{1/2}$,
\begin{equation}
\sigma ^{2}=\frac{1}{2}\sigma ^{ij}\sigma _{ij},\hspace{0.4cm}\sigma
^{2}\geq 0,  \label{shear2}
\end{equation}
where $\sigma ^{jl}=\delta ^{ij}\delta ^{kl}\sigma _{ik}$. One can easily
prove that $\sigma =0\ $if and only if $\sigma _{ij}=0$.

The scalar of expansion $\theta $ determines the change of the volume of a
fluid particle keeping its (spherical) form unchanged. It is useful to
define a representative scalar $R=R(x^{k},t)$ of a fluid particle by the
equation
\begin{equation}
\frac{1}{R}\frac{dR}{dt}=\frac{1}{3}\theta .  \label{length}
\end{equation}
If the expansion scalar vanishes everywhere (\ref{incompressible}), the
fluid is incompressible. If the expansion scalar (\ref{expansion}) does not
vanish identically for a fluid distribution (\ref{compressible}), the fluid
is compressible.

The magnitude of vorticity $\omega $ is determined by $\omega =\mid \omega
^{2}\mid ^{1/2}$,
\begin{equation}
\omega ^{2}=\frac{1}{2}\omega ^{ij}\omega _{ij},\hspace{0.4cm}\omega
^{2}\geq 0,  \label{vorticity2}
\end{equation}
where $\omega ^{jl}=\delta ^{ij}\delta ^{kl}\omega _{ik}$ and it is easy to
show that $\omega ^{2}=\omega ^{i}\omega _{i}$. One can prove that $\omega
=0 $ if and only if\ either $\omega _{ij}=0$ or $\omega ^{i}=0$.

\section{The Conservation of Energy}

\label{*ce}

On the basis of the Principles of conservation of mass (\ref{mass-reg}) and
linear momentum (\ref{momentum-conservation}) one can establish the transfer
equation \cite{Serr:1959}, \cite{Aris:1962} for the kinetic energy of an
arbitrary moving fluid region $\Sigma (t)$
\begin{equation}
\mathcal{K}(t)=\frac{1}{2}\int\limits_{\Sigma (t)}\rho \delta
_{ij}u^{i}u^{j}dV.  \label{kinetic}
\end{equation}

\begin{theorem}[The energy transfer equation]
For any region $\Sigma (t)$ following the particle paths of the fluid which
conserves mass (\ref{mass-reg}) and linear momentum (\ref
{momentum-conservation}) the rate of change of its kinetic energy (\ref
{kinetic}) is equal to the rate at which work is being done on the region by
external forces reduced by the work which is being done to change the volume
and the shape of the region and by the work which is being done to rotate
the region\textsc{\ }
\begin{equation}
\frac{d}{dt}\mathcal{K}(t)=\int\limits_{\Sigma (t)}\rho \delta
_{ij}u^{i}F^{j}dV+\oint\limits_{\partial \Sigma (t)}\delta
_{ij}u^{i}s_{(n)}^{j}dS-\int\limits_{\Sigma (t)}\theta
_{ij}T^{(ij)}dV-\int\limits_{\Sigma (t)}\omega _{ij}T^{[ij]}dV.
\label{energy-transfer}
\end{equation}
\end{theorem}

\noindent \textbf{Proof.}\hspace{0.4cm}The energy transfer equation (\ref
{energy-transfer}) can be easily derived by using (\ref{eq-continuity-gen}),
the Cauchy equation of motion (\ref{eqs-motion-cauchy}) and a decomposition
of the fluid stress tensor $T^{ij}$ into its symmetric and antisymmetric
parts to produce the last term in accordance with the definitions of the
fluid deformation tensor $\theta _{ij}$ (\ref{decomposition}) and vorticity
tensor $\omega _{ij}$ (\ref{vorticity}).\hspace{0.4cm}\textbf{QED}\vspace{
0.2cm}

The last two terms involving the interaction of the symmetric and
antisymmetric parts of the fluid stress tensor with the deformation and
vorticity tensors can be considered as a \textquotedblleft dissipation"
part. This part of the power in (\ref{energy-transfer}) may well be
recoverable, but the rest must be accounted for as heat.

Considering now a class of chemically inert fluids without diffusion the
total energy $\mathcal{T}(t)$ of an arbitrary moving fluid region $\Sigma
(t) $ is defined as a sum of its kinetic energy $\mathcal{K}(t)$ (\ref
{kinetic}) and its internal energy $\mathcal{E}(t)$,
\begin{equation}
\mathcal{T}(t)=\mathcal{K}(t)+\mathcal{E}(t)=\frac{1}{2}\int\limits_{\Sigma
(t)}\rho \delta _{ij}u^{i}u^{j}dV+\int\limits_{\Sigma (t)}\rho EdV,
\label{energy-total}
\end{equation}
where $E=E(x^{i},t)$ is the specific internal energy. The thermodynamical
interpretation of $E$ is different for the compressible and incompressible
fluids, (\ref{compressible}) and (\ref{incompressible}) correspondingly.
Indeed, for the compressible fluid its density is a function, $\rho
(x^{i},t)\neq \mathrm{const}$, due to the continuity equation (\ref
{eq-continuity}). Then it is naturally to consider the specific internal
energy $E$ as a thermodynamical state variable satisfying\vspace{0.2cm}

\noindent \textbf{The Postulate of thermodynamical state for the
compressible fluids}
\begin{equation}
TdS=dE+pd\tau ,\hspace{0.4cm}\tau =\frac{1}{\rho }.  \label{thermo-state}
\end{equation}
\vspace{0.2cm}

\noindent Here $T$ is the absolute temperature, $S$ is the specific entropy, $p$ is
the pressure, and $\tau $ is the specific fluid volume. Due to this equation
of the thermodynamic state (\ref{thermo-state}) the fluid pressure $p$ is a
thermodynamical variable and there exists in general an equation of state
\begin{equation}
p=p(\rho ,S)\hspace{0.4cm}\mathrm{or}\hspace{0.4cm}\rho =\rho (p,T).
\label{eq-state-thermo}
\end{equation}
The equation of state is usually chosen or derived on the basis of a
phenomenological macroscopic and/or a microscopic physical model of the
fluid substance.

For the incompressible fluid $\rho (x^{p},t)=\mathrm{const}$ due to the
continuity equation (\ref{eq-continuity}) and the pressure cannot be a
thermodynamical variable. In this case the hydrodynamic pressure is not
defined as a function of the state of the fluid. Taking one or another
definition of pressure here, one should remember that it may not be
necessarily the physical pressure measured in experiments. The freedom in
choice of the pressure is usually utilized to simplify the equations of
motion.\vspace{0.2cm}

\noindent \textbf{The Postulate of thermodynamical state for the
incompressible fluids}
\begin{equation}
TdS=dE,\hspace{0.4cm}\rho \equiv \mathrm{const}.  \label{thermo-state-incomp}
\end{equation}
\vspace{0.2cm}

In general, due to possible internal dissipation processes a fluid can allow
a heat transfer which is described by a heat flux vector $
q^{i}=q^{i}(x^{j},t)$ with the physical dimension of energy per unit area
per unit time. The fundamental postulate of conservation of total energy
reads as follows \cite{Serr:1959}, \cite{True-Toup:1960}, \cite{Aris:1962}.
\vspace{0.2cm}

\noindent \textbf{The First law of thermodynamics for fluids (The Postulate
of conservation of total energy)}\hspace{0.2cm}\emph{The rate of change of
total energy }$\mathcal{T}(t)$\emph{\ (\ref{energy-total}) of an arbitrary
fluid region }$\Sigma (t)$\emph{\ is equal to the rate at which work is
being done on the region plus the rate at which heat is conducted into the
region}
\begin{equation}
\frac{d}{dt}\left[ \mathcal{K}(t)+\mathcal{E}(t)\right] =\int\limits_{\Sigma
(t)}\rho \delta _{ij}u^{i}F^{j}dV+\oint\limits_{\partial \Sigma (t)}\delta
_{ij}u^{i}s_{(n)}^{j}dS-\oint\limits_{\partial \Sigma (t)}q^{i}dS_{i}.
\label{energy-total-post}
\end{equation}
\vspace{0.2cm}

The following fundamental theorem establishing the total energy equation now
can be proved.

\begin{theorem}[The total energy equation]
The total energy equation
\begin{equation}
\rho \frac{dE}{dt}=\theta _{ij}T^{(ij)}+\omega _{ij}T^{[ij]}-\frac{\partial
q^{i}}{\partial x^{i}}  \label{eq-total-energy}
\end{equation}
is necessary and sufficient for the First law of thermodynamics for fluids
(\ref{energy-total-post}) to be satisfied for any region which follows the
fluid particle paths as that the energy transfer equation (\ref
{energy-transfer}) for the rate of change of its kinetic energy holds.
\end{theorem}

\noindent \textbf{Proof.}\hspace{0.4cm}Using the definition of the total
energy (\ref{energy-total}) contained in an arbitrary fluid region $\Sigma
(t)$, Eq. (\ref{eq-continuity-gen}), the Gauss theorem and the kinetic
energy transfer equation (\ref{energy-transfer}), the First law of
thermodynamics for fluids (\ref{energy-total-post}) takes the form

\begin{equation}
\frac{d}{dt}\mathcal{E}(t)=\int\limits_{\Sigma (t)}\rho \frac{dE}{dt}
dV=\int\limits_{\Sigma (t)}\left( \theta _{ij}T^{(ij)}+\omega _{ij}T^{[ij]}-
\frac{\partial q^{i}}{\partial x^{i}}\right) dV.  \label{energy-total-post2}
\end{equation}
Now the necessity of the total energy equation (\ref{eq-total-energy}) to
satisfy Eq. (\ref{energy-total-post2}) is evident. The sufficiency can be
easily seen from (\ref{energy-total-post2}) as a result of arbitrariness of
the moving fluid region $\Sigma (t).$\hspace{0.4cm}\textbf{QED}\vspace{0.2cm}

The total energy (\ref{eq-total-energy}) in the case of a symmetric fluid
stress tensor, $T^{[ij]}=0$, is due to Neumann \cite{Neum:1894}. Further
assumptions on the form of the fluid stress tensor $T^{ij}$, the heat flux
vector $q^{i}$ and the equation of state (\ref{eq-state-thermo}) relating
the thermodynamical variables should be made to apply the First law of
thermodynamics for fluids (\ref{energy-total-post}) for the resolution of
the energy balance for a particular moving fluid.

The fundamental postulate of the production of entropy can be formulated for
moving fluids in the following form \cite{Serr:1959}, \cite{True-Toup:1960},
\cite{Aris:1962}. \vspace{0.2cm}

\noindent \textbf{The Second law of thermodynamics for fluids (The Postulate
of increasing entropy)}\hspace{0.2cm}\emph{The rate of increase of entropy
of an arbitrary fluid region }$\Sigma (t)$\emph{\ should not be less than
the heat conducted into the region divided by the temperature at which the
heat transfer occurs}
\begin{equation}
\frac{d}{dt}\int\limits_{\Sigma (t)}\rho SdV\geqq -\oint\limits_{\partial
\Sigma (t)}\frac{1}{T}q^{i}dS_{i}.  \label{entropy-increase}
\end{equation}
\vspace{0.2cm}

\begin{theorem}[The local increase of entropy]
The following inequality for the local increase of entropy
\begin{equation}
\rho \frac{dS}{dt}\geqq -\frac{1}{T}\frac{\partial q^{i}}{\partial x^{i}}+
\frac{q^{i}}{T^{2}}\frac{\partial T}{\partial x^{i}}
\label{entropy-increase2}
\end{equation}
is necessary and sufficient for the Second law of thermodynamics for fluids
(\ref{entropy-increase}) to be satisfied for any region which follows the
fluid particle paths with the conservation of its mass (\ref{mass-reg}).
\end{theorem}

\noindent \textbf{Proof.}\hspace{0.4cm}Using (\ref{eq-continuity-gen}) and
the Gauss theorem, the Second law of thermodynamics for fluids (\ref
{entropy-increase}) takes the form

\begin{equation}
\int\limits_{\Sigma (t)}\left( \rho \frac{dS}{dt}+\frac{1}{T}\frac{\partial
q^{i}}{\partial x^{i}}-\frac{q^{i}}{T^{2}}\frac{\partial T}{\partial x^{i}}
\right) dV\geqq 0.  \label{entropy-increase3}
\end{equation}
Now the necessity of the inequality (\ref{entropy-increase2}) for the local
increase of entropy (\ref{entropy-increase}) to ensure the integral
inequality (\ref{entropy-increase3}) follows because an integral of a
function everywhere nonnegative is nonnegative. The sufficiency can be
easily seen from (\ref{entropy-increase3}) as a result of arbitrariness of a
moving fluid region $\Sigma (t)$, which forces the integrand to be
nonnegative everywhere$.$\hspace{0.4cm}\textbf{QED}\vspace{0.2cm}

\section{The Fluid Stress Tensor and the Conservation of Angular Momentum}

\label{*fstcam}

In general, the second rank tensor $T^{ij}(x^{k},t)$ has no algebraic
symmetry in its indices. From a physical point of view that means that a
fluid can be generally subjected to extraneous torques due to action of the
external forces $F^{i}(x^{j},t)$. Such fluids are called polar and they are
capable of transmitting extraneous stress couples and carrying the body
torques \cite{Serr:1959}, \cite{Aris:1962}, \cite{Hame:1927}. If a fluid can
only admit the torques arising from the moments of direct forces, it is
called nonpolar. Thus, certain assumptions concerning the nature of the
forces exerted across fluid surface elements will specify the properties of
the stress tensor. Let us accept the following postulate \cite{Serr:1959},
\cite{Hame:1927} about the fluid stress tensor.\vspace{0.2cm}

\noindent \textbf{The Boltzmann Postulate}\emph{\hspace{0.2cm}The fluid
stress tensor }$T^{ij}=T^{ij}(xk,t)$ \emph{is symmetric}
\begin{equation}
T^{ij}=T^{ji}.  \label{bolzmann}
\end{equation}
\vspace{0.2cm}

One can prove an important theorem \cite{Serr:1959}, \cite{Aris:1962}, \cite
{Cauc:1827} that fluids with a symmetric stress tensor $T^{ij}(x^{k},t)$
(\ref{bolzmann}) conserve the angular momentum $L_{i}(t)$ of an arbitrary
fluid region $\Sigma (t)$
\begin{equation}
L_{i}(t)=\int\limits_{\Sigma (t)}\rho \varepsilon _{ijk}x^{j}u^{k}dV.
\label{ang-momentum}
\end{equation}

\noindent \textbf{The Principle of conservation of angular momentum}\emph{
\hspace{0.2cm}The rate of change of angular momentum }$L_{i}(t)$\emph{\ (\ref
{ang-momentum}) of an arbitrary fluid region }$\Sigma (t)$\emph{\ is equal
to the torque due to the total resultant force acting on the fluid volume
within the region}\textsc{\ }
\begin{equation}
\frac{d}{dt}L_{i}(t)=\frac{d}{dt}\int\limits_{\Sigma (t)}\rho \varepsilon
_{ijk}x^{j}u^{k}dV=\int\limits_{\Sigma (t)}\rho \varepsilon
_{ijk}x^{j}F^{k}dV+\oint\limits_{\partial \Sigma (t)}\varepsilon
_{ijk}x^{j}s_{(n)}^{k}dS.  \label{ang-momentum-conservation}
\end{equation}
\vspace{0.2cm}

\begin{theorem}[The conservation of angular momentum]
For any fluid moving with conservation of mass (\ref{mass-reg}) and linear
momentum (\ref{momentum-conservation}) the Boltzmann postulate for the fluid
stress tensor (\ref{bolzmann}) is the necessary and sufficient condition for
the fluid to conserve also the angular momentum (\ref
{ang-momentum-conservation}) for any region following the fluid particle
paths.
\end{theorem}

\noindent \textbf{Proof.}\hspace{0.4cm}By using (\ref{eq-continuity-gen})
and the Cauchy equation of motion (\ref{eqs-motion-cauchy}) one can show
that the rate of change of the angular momentum (\ref{ang-momentum}) can be
expressed in the following form:
\begin{align}
\frac{d}{dt}L_{i}(t)& =\frac{d}{dt}\int\limits_{\Sigma (t)}\rho \varepsilon
_{ijk}x^{j}u^{k}dV=\int\limits_{\Sigma (t)}\rho \varepsilon _{ijk}x^{j}\frac{
d}{dt}u^{k}dV=  \notag \\
& \int\limits_{\Sigma (t)}\rho \varepsilon
_{ijk}x^{j}F^{k}dV+\oint\limits_{\partial \Sigma (t)}\varepsilon
_{ijk}x^{j}s_{(n)}^{k}dS-\int\limits_{\Sigma (t)}\varepsilon _{ijk}T^{jk}dV.
\label{t-symmetric}
\end{align}
Since $\varepsilon _{ijk}T^{jk}(x^{l},t)\equiv 0$ if the fluid stress tensor
is symmetric, the Boltzmann postulate (\ref{bolzmann}) necessarily brings
the Principle of conservation of angular momentum (\ref
{ang-momentum-conservation}). Sufficiency of the Boltzmann postulate follows
from (\ref{ang-momentum-conservation}) and (\ref{t-symmetric}) due to
arbitrariness of the fluid region $\Sigma (t)$ when the vanishing of the
last integral in (\ref{t-symmetric}) requires $\varepsilon
_{ijk}T^{jk}(x^{j},t)\equiv 0$ everywhere.\hspace{0.4cm}\textbf{QED}\vspace{
0.2cm}

The interpretation of the components of a symmetric fluid stress tensor is
well-known (see for example, \cite{Serr:1959}, \cite{True-Toup:1960}, \cite
{Aris:1962}).

\section{The Navier-Stokes Equation and Its Physical Status}

\label{*nseps}

The different physical hypotheses about the nature of moving fluids and
forces causing their motion bring different analytical forms for the fluid
stress tensor. As a result, historically various forms of the Cauchy
equation of motion (\ref{eq1-cauchy}) have been discovered to describe the
dynamics of various fluids. By restriction to the class of fluids without
heat transfer,
\begin{equation}
q^{i}(x^{j},t)=0,\hspace{0.4cm}T=\mathrm{const},  \label{no-heat}
\end{equation}
there are three forms of the equations of motion of interest and relevance
here.\vspace{0.2cm}

\noindent \textbf{The Euler equation of motion for the perfect fluid }$
T^{ij}=-\delta ^{ij}p$
\begin{equation}
\rho \left( \frac{\partial u^{i}}{\partial t}+u^{j}\frac{\partial u^{i}}{
\partial x^{j}}\right) =\rho F^{i}-\delta ^{ij}\frac{\partial p}{\partial
x^{j}}  \label{eq-euler}
\end{equation}
\vspace{0.2cm}

\noindent \cite{Serr:1959}, \cite{Aris:1962}, \cite{Eule:1755}. Here $p=p(x^{j},t)$ is
the fluid pressure which depends on the thermodynamical state of fluid. It
can be positive, $p>0$, which tends to compress a fluid particle, and
negative, $p\,<0$, which results in an expansion of a fluid particle. To
close the system of the four first order partial differential equations
(\ref{eq2-continuity}) and (\ref{eq-euler}) for four unknown functions of the
components of the fluid velocity $u^{i}(x^{j},t)$ and the density $\rho
(x^{i},t)$ provided the external forces $F^{i}(x^{j},t)$ and the initial and
boundary conditions for the velocity and density are given, one must specify
in addition to the system the of state (\ref{eq-state-thermo}) which does
not depend on temperature due to (\ref{no-heat})\vspace{0.2cm}

\noindent \textbf{The equation of state }
\begin{equation}
p=p(\rho )\hspace{0.4cm}\mathrm{or}\hspace{0.4cm}\rho =\rho (p).
\label{eq-state}
\end{equation}
\vspace{0.2cm}

\noindent \textbf{The Stokes equation of motion for the Stokean fluid }$
T^{ij}=(-p+\alpha )\delta ^{ij}+\beta \theta ^{ij}+\gamma \delta _{kl}\theta
^{ik}\theta ^{jl}$
\begin{equation}
\rho \left( \frac{\partial u^{i}}{\partial t}+u^{j}\frac{\partial u^{i}}{
\partial x^{j}}\right) =\rho F^{i}-\delta ^{ij}\frac{\partial p}{\partial
x^{j}}+\frac{\partial }{\partial x^{j}}\left( \beta \theta ^{ij}\right)
+\delta _{kl}\frac{\partial }{\partial x^{j}}\left( \gamma \theta
^{ik}\theta ^{jl}\right)   \label{eq-stokes}
\end{equation}
\vspace{0.2cm}

\noindent \cite{Serr:1959}, \cite{Stok:1845}, \cite{Aris:1962}. Here $\alpha =\alpha
(x^{i},t)$, $\beta =\beta (x^{i},t)$ and $\gamma =\gamma (x^{i},t)$ are the
scalar functions which depend on the thermodynamical state of fluid as well
as on the principal invariants of the symmetric deformation tensor $\theta
^{ij}=\theta ^{ij}(x^{k},t)$, $\theta ^{ij}=\delta ^{ik}\delta ^{jl}\theta
_{kl}$ (\ref{deformation}). They describe definite types of viscous response
of the fluid. For example, a choice of $p$, $\alpha $, $\beta $ and $\gamma $
so that $T^{ij}$ is linear in $\theta ^{ij}$ provides the classical
Cauchy-Poisson law of viscosity. The Stokean fluid is the most general type
of a nonelastic fluid. To close the system of Eqs. (\ref{eq2-continuity}),
(\ref{eq-state}) and (\ref{eq-stokes}) for four unknown functions of the
components of the fluid velocity $u^{i}(x^{j},t)$ and density $\rho (x^{i},t)
$ provided the external forces $F^{i}(x^{j},t)$ and the initial and boundary
conditions for the velocity and the density are given, one must specify in
addition the viscosity functions $\alpha (x^{i},t)$, $\beta (x^{i},t)$ and $
\gamma (x^{i},t)$.\vspace{0.2cm}

\noindent \textbf{The Navier-Stokes equation of motion of classical
hydrodynamics for the Newtonian fluid} $T^{ij}=\delta ^{ij}(\lambda \theta
-p)+2\mu \theta ^{ij}$
\begin{equation}
\rho \left( \frac{\partial u^{i}}{\partial t}+u^{j}\frac{\partial u^{i}}{
\partial x^{j}}\right) =\rho F^{i}-\delta ^{ij}\frac{\partial p}{\partial
x^{j}}+\delta ^{ij}\frac{\partial }{\partial x^{j}}\left( \lambda \theta
\right) +2\frac{\partial }{\partial x^{j}}\left( \mu \theta ^{ij}\right)
\label{eq-navier-stokes}
\end{equation}
\vspace{0.2cm}

\noindent \cite{Serr:1959}, \cite{Navi:1827}, \cite{Stok:1845}, \cite{Aris:1962}. Here
$\lambda =\lambda (x^{i},t)$ and $\mu =\mu (x^{i},t)$ are the scalar
functions of the first and second coefficients of viscosity. For the case of
constant viscosity coefficient, $\lambda (x^{i},t)=\mathrm{const}$ and $\mu
(x^{i},t)=\mathrm{const}$, the Navier-Stokes equations becomes much simpler
\begin{equation}
\rho \left( \frac{\partial u^{i}}{\partial t}+u^{j}\frac{\partial u^{i}}{
\partial x^{j}}\right) =\rho F^{i}-\delta ^{ij}\frac{\partial p}{\partial
x^{j}}+\left( \lambda +\mu \right) \delta ^{ij}\frac{\partial ^{2}u^{k}}{
\partial x^{j}\partial x^{k}}+\mu \delta ^{kl}\frac{\partial ^{2}u^{i}}{
\partial x^{k}\partial x^{l}}.  \label{eq-navier-stokes2}
\end{equation}
Its further simplification for the case of incompressible fluid (\ref
{incompressible}) has been especially thoroughly studied in connection with
the phenomenon of fluid and gas turbulence \cite{Moni-Yagl:1971}, \cite
{Moni-Yagl:1975}-\cite{Stan:1985}, \cite{FMRT:2001}, \cite{Batc:1953}.

The Newtonian fluid is a linear Stokean fluid with the Cauchy-Poisson law of
viscosity. For a Newtonian fluid its stress tensor depends linearly on the
rate of fluid deformation. It is generally believed that the Newtonian fluid
is an adequate physical model when the fluid deformations are relatively
small in comparison with, say, the ratio of some reference speed and
reference length. However, this physical hypothesis is not well understood
as yet \cite{Serr:1959}. It is not to be derived from experiments, nor can
it be proved by abstract modelling. The Navier-Stokes equation (\ref
{eq-navier-stokes}) is the dynamical equation of classical hydrodynamics.
Since their discovery they successfully describe various regimes of fluid
motion including turbulence for a wide class of physical gaseous and liquid
substances.

A closed system of the equations of classical hydrodynamics without heat
transfer includes Eqs. (\ref{eq2-continuity}), (\ref{eq-state}) and (\ref
{eq-navier-stokes2}) for four unknown functions of the components of the
fluid velocity $u^{i}(x^{j},t)$ and the fluid density $\rho (x^{i},t)$
provided the external forces $F^{i}(x^{j},t)$, the viscosity coefficient
functions $\lambda (x^{i},t)$ and $\mu (x^{i},t)$ and the initial and
boundary conditions for the velocity and the density are given.

The mathematical difficulties in solving the Navier-Stokes equation are well
known, see \cite{FMRT:2001} for a discussion and further references. Because
the 3-dimensional Navier-Stokes equation is not a well-posed partial
differential system, no theorem of existence of smooth, physically
reasonable solutions is not known as yet. These problems originate in the
nonlinear structure of the Euler equation (\ref{eq-euler}) and the
Navier-Stokes equation (\ref{eq-navier-stokes2})\ due to the presence of the
so-called inertial term $u^{j}\partial u^{i}/\partial x^{j}$. These
equations are among the very few equations of mathematical physics for which
the nonlinearity arises not from the physical attributes of the system, such
as a physical interaction, but rather from the kinematical aspect of fluid
motion. Another source of the difficulties for the Navier-Stokes equation is
due to the viscosity term which brings the second order operator into the
first order equations, which involves the singular perturbations.

As far as the physical status of the Navier-Stokes equation is concerned it
must be regarded as the macroscopic equation to be able to describe the
dynamics of fluids considered as continuous physical substances. The flow of
fluids at the microscopic level is governed by the statistical mechanics of
fluids. The appropriate statistics is given by the solution of the Boltzmann
equation which represents the evolution of the distribution function in
dependence on positions and velocities of colliding particles as a result of
thermal excitation. The collisions are described by the so-called collision
integral which, in principle, contains all information about interaction
between particles and it forms the collision operator. One must evaluate
this operator to obtain a set of macroscopic equations for one or another
physical configuration of interacting (colliding) particles. Only for the
case of dilute gases when one is limited to the case of the evolution of a
single particle distribution with only binary collisions taken into account
it has been possible to find a satisfactory solution given by Chapman \cite
{Chap:1916} and Enskog \cite{Ensk:1917}. It has been shown that the
Boltzmann equation with this collision operator gives the equation of the
conservation of mass and the equation of the conservation of linear momentum
which is the Navier-Stokes equation for an incompressible fluid \cite
{Uhle-Ford:1963}. Although such a derivation has been carried for dilute
gases, a derivation for liquid fluids when the binary collisions are almost
of negligible significance and all liquid physics is due to collisions
between particle clusters is still an open problem. In this situation the
Navier-Stokes equation (\ref{eq-navier-stokes2}) is taken as a reasonable
model for liquid flows at least for the simple constant viscosity.

\section*{Acknowledgments}

I would like to thank Alan Coley for his hospitality in Dalhousie
University. The work has been supported in part by the Swiss National
Science Foundation, Grant 7BYPJ065731.

\end{document}